\documentclass[12pt]{article}
\usepackage{amsmath}
\usepackage{mathptmx}
\usepackage{amsthm}
\usepackage{amssymb}
\usepackage{cancel}
\usepackage{verbatim}

\usepackage{geometry}
\usepackage{filecontents}

\usepackage{xcolor}

\usepackage{hyperref}
\hypersetup{
	colorlinks,
	citecolor=blue,
	filecolor=blue,
	linkcolor=blue,
	urlcolor=blue
}

\usepackage[compress,numbers]{natbib}

\usepackage[affil-it]{authblk}

\title{Charged Boundary States in the Schwinger Model}
\author{Adar Sharon}
\author{Gilad Weiss}
\affil{Weizmann Institute of Science, Rehovot, Israel.}
\date{\vspace{-5ex}}

\begin{document}

\maketitle

\begin{abstract}
	QED in 1+1 dimensions possesses two rare and interesting properties - It is both exactly solvable and confining. The combination of these two properties makes it the perfect candidate for a toy model for QCD. We study this model on an interval, where new features of the theory are revealed. We show that on an interval, the model admits charged states, which is unexpected for a confining theory. We show that the charged states are boundary states, and calculate their mass. This result could lead one to expect a similar result in QCD, where single quarks could be isolated near a boundary.
\end{abstract}

\section{Introduction}
Quantum Electrodynamics in $ 1+1 $ dimensions, also called the Schwinger Model, has proven to be a rich and intriguing theory. The model was studied extensively  \cite{Schwinger1,Schwinger2,ColemanJackiwSusskind,Coleman1976,Johnson1963,Manton:1985} due to its many interesting aspects, ranging from an axial anomaly \cite{Johnson1963,Manton:1985} to a mass generating mechanism for a gauge field \cite{Schwinger1,Schwinger2}. Two features of this theory stand out: first, it is confining, and second, its spectrum can be solved exactly. The combination of these two properties make the Schwinger Model a perfect toy model for Quantum Chromodynamics (QCD).

One might be curious as to what happens to the Schwinger Model when a boundary is introduced. Theories with boundaries have been gaining increased attention, partly due to their many applications in Condensed Matter physics. One important aspect of theories with boundaries is the appearance of boundary states, which often display interesting features. One might then wonder whether boundary states also appear in a theory which is confining, like the Schwinger Model. In particular, one might ask whether a confining theory, whose spectrum in free space consists only of neutral excitations, will admit charged edge states. Intuitively, this could be possible for a theory in which confinement was a result of the existence of flux tubes. Assuming flux tubes may end at a boundary, one can imagine a flux tube connecting a charged particle to the boundary, thus giving rise to a charged boundary state. This situation is certainly plausible in the Schwinger Model, and might also be possible in QCD.

This paper focuses on finding charged boundary states in the Schwinger Model. We begin by reviewing the method of abelian bosonization in section \ref{bosonization}, and proceed by applying this method to the Schwinger Model in  section \ref{bos_of_schwinger}. In section \ref{RxR} we review the solution of the Schwinger Model on $ \mathbb{R}\times\mathbb{R} $, and in section \ref{S1} we review the solution of the Schwinger Model on $ \mathbb{R} \times S^1  $, taking special care of the zero modes. Finally, in section \ref{RxI} we study the Schwinger Model on $ \mathbb{R}\times \left[0,L\right] $. We find charged states which are confined to the boundary for both Dirichlet and Neumann boundary conditions for the gauge field, and find their energy. Specifically, in the limit $e L\rightarrow\infty $ we find the energy of $ k $ electrons confined to the boundary to be $ E_0=\frac{e\sqrt{\pi}}{2}k^2 $.

\section{Bosonization}\label{bosonization}

We review the method of abelian bosonization. Since spin does not exist in $ 1+1 $ dimensions, we might expect the existence of a duality between a free fermion and a free boson. This duality was named bosonization, and was first introduced by Coleman \cite{Coleman1975} and Mandelstam \cite{Mandelstam1975}. We begin by reviewing the free boson and free fermion, and proceed to obtain the duality between them.
	
The theory of a free massless boson in $ 1+1$ dimensions is defined by the action	
\begin{equation}
	S=\frac{1}{2}\int d^{2}x \left(\partial_{\mu}\varphi \right)^2
	\label{eq:boson_action}
\end{equation}
and contains two classically conserved currents. The first is associated with translation symmetry of the boson $ \varphi $, and is given by  $ k_{\mu}=\partial_{\mu}\varphi $. The second is unique to $ 1+1 $ dimensions, and is given by $ \widetilde{k}_{\mu}=\epsilon_{\mu\nu}k^{\nu}_{\varphi}=\partial_{\mu}\widetilde{\varphi}$ (where $\widetilde{\varphi}$ is the dual field, defined by $\partial_{\mu}\widetilde{\varphi}=\epsilon_{\mu\nu}\partial^{\nu}\varphi$).

The theory of a free massless fermion is defined by the action
\begin{equation}
	S
	=i\int d^{2}x \bar{\psi}\cancel{\partial}\psi
\end{equation}
and also contains two classically conserved currents. The first is the vector current, given by $ j_{\mu}=\bar{\psi}\gamma_{\mu}\psi $, and the second is the axial current, given by $ \widetilde{j}_\mu=\bar{\psi}\gamma_\mu\gamma_5\psi $. In $ 1+1 $ dimensions they are related by $ j_{\mu}=\epsilon_{\mu\nu}\widetilde{j}^{\nu} $, which justifies the choice of notation for $ \widetilde{j}_{\mu} $. 
	
Any duality between the two theories must also map each symmetry of one theory to a corresponding symmetry in the other. Assuming we choose to map $ k_\mu $ to $ \widetilde{j}_\mu $ and $ \widetilde{k}_\mu $ to $ j_\mu $, we have the following form for the duality:
\begin{equation}
	\psi=\left(\begin{array}{c}
	\psi_{L}\\
	\psi_{R}
	\end{array}\right)\longleftrightarrow\left(\begin{array}{c}
	:e^{i\lambda\left(\varphi-\widetilde{\varphi}\right)}:\\	:e^{-i\lambda\left(\varphi+\widetilde{\varphi}\right)}:
	\end{array}\right)\label{eq:pre_bosonization}
\end{equation}
where $ :A: $ denotes the normal ordering of the operator $ A $. Indeed, a translation of $ \varphi $ in equation \eqref{eq:pre_bosonization} corresponds to an axial transformation. 

The constant $ \lambda $ in equation \eqref{eq:pre_bosonization} is obtained by demanding that the right-hand side of equation \eqref{eq:pre_bosonization} obeys fermionic particle statistics \cite{Senechal}. The result is $ \lambda=\sqrt{\pi} $, and we have obtained the required duality:
\begin{subequations}\label{eq:bosonization}
	\begin{align}
		\psi=\left(\begin{array}{c}
		\psi_{L}\\
		\psi_{R}
		\end{array}\right) \quad&\longleftrightarrow\quad \left(\begin{array}{c}
		:e^{i\sqrt\pi\left(\varphi-\tilde{\varphi}\right)}:\\
		:e^{-i\sqrt\pi\left(\varphi+\tilde{\varphi}\right)}:
		\end{array}\right)
		\label{eq:bosonization_1}\\
		j_\mu \quad&\longleftrightarrow\quad -\frac{1}{\sqrt{\pi}}\widetilde{k}_\mu
		\label{eq:bosonization_2}\\
		\widetilde{j}_\mu \quad&\longleftrightarrow\quad -\frac{1}{\sqrt{\pi}}k_\mu
		\label{eq:bosonization_3}
	\end{align}
\end{subequations}

We now comment on several subtleties of the duality. First, we note that the boson  $ \varphi $ appearing in the duality is compact with radius $ R_{\varphi,free}=2\sqrt{\pi} $, so that $ \varphi\sim\varphi+2\sqrt{\pi} $. Indeed, translating $ \varphi $ by $ 2\sqrt{\pi} $ will keep the exponent in \eqref{eq:bosonization_1} invariant\footnote{Additionally, note that sending $ \varphi $ to $\varphi + \sqrt{\pi}$ simply sends $ \psi $ to $ -\psi $.}.

Naively, this leads us to expect the global symmetry group of the free boson to be the product group $ U_T\left(1\right)\times U_{\widetilde{T}} \left(1\right) $, representing independent translations in $ \varphi$ and $ \widetilde{\varphi} $ by $ \alpha,\beta $ respectively for $ 0\leq\alpha,\beta\leq2\sqrt{\pi} $. The correct result is more subtle, and is more easily understood in the fermionic theory. For a free massless fermion, the correct global symmetry group is $ \left(U_A\left(1\right)\times U_V\left(1\right)\right)/\mathbb{Z}_2 $, as opposed to the more commonly mentioned $ U_A\left(1\right)\times U_V\left(1\right) $, representing independent axial and vector transformations. This results from the fact that for the free fermion, the transformations resulting from the group elements $ \left(e^{i\alpha \gamma_5},e^{i\beta}\right) $ and $ \left(-e^{i\alpha \gamma_5},-e^{i\beta}\right) $ are identical. Thus, for example, we must identify the group element $ (-1,-1) $ with the identity $(1,1)$. In the bosonic theory, the duality \eqref{eq:bosonization_1} then results in the translations by $ \left(\alpha,\beta\right) $ being identified with the translations by $ \left(\alpha+\sqrt{\pi},\beta+\sqrt{\pi}\right) $, meaning that the correct global symmetry group of the resulting boson is thus $ \left(U_T\left(1\right)\times U_{\widetilde{T}} \left(1\right)\right)/\mathbb{Z}_2 $.

We have thus found that the free massless fermion is dual to the free massless (and compact) boson. The method of bosonization can be used to find dualities between more complex theories of fermions and bosons in $ 1+1 $ dimensions. For instance, the Thirring model is also dual to the free massless boson, while a massive fermion is dual to the Sine-Gordon model \cite{Coleman1975}.

\section{The Schwinger Model}\label{bos_of_schwinger}

\subsection{Bosonization of the Schwinger Model}

We now apply the method of abelian bosonization to the Schwinger Model, following \cite{Coleman1976}.
Consider the action for the Schwinger Model:
\begin{equation}
S=\int d^{2}x\left[-\frac{1}{4e^{2}}F_{\mu\nu}^{2}+i\overline{\psi}\cancel{\partial}\psi-j_{\mu}A^{\mu}\right]\label{eq:SMAction}
\end{equation}
which is just the action for QED in $ 1+1 $ dimensions. The current $ j_{\mu} $ is the vector current defined in section \ref{bosonization}, and $ F_{\mu\nu} = \partial_{\mu}A_{\nu} - \partial_{\nu}A_{\mu} $ is the field strength tensor. 

As shown in section \ref{bosonization}, a free massless fermion is dual to a free massless boson, and so we replace the kinetic term for the fermion $ \psi $ with that of a boson $ \varphi $. 
Additionally, the duality \eqref{eq:bosonization} allows us to replace $ j_{\mu} $ with its bosonized counterpart $ -\frac{1}{\sqrt{\pi}}\widetilde{k}_{\mu} $.
These replacements result in the following action:
\begin{equation}
S=\int d^{2}x\left[-\frac{1}{4e^{2}}F_{\mu\nu}^{2}+
\frac{1}{2}\left(\partial_{\mu}\varphi\right)^{2}+
\frac{1}{\sqrt{\pi}}\epsilon_{\mu\nu} A^{\mu} \partial^{\nu}\varphi \right]
\label{eq:temp_bos_action}
\end{equation}
after an integration by parts on the last term in \eqref{eq:temp_bos_action} and completing the square, one obtains a more compact expression:
\begin{equation}
S=\int d^{2}x\left[\frac{1}{2e^{2}}\left(F_{01}-\frac{e^{2}}{\sqrt{\pi}}\varphi\right)^{2}+\frac{1}{2}\left(\partial_{\mu}\varphi\right)^{2}-\frac{1}{2}\frac{e^{2}}{\pi}\varphi^{2}\right]
\label{eq:bos_action}
\end{equation}
we shall call equation \eqref{eq:bos_action} the bosonized form of the Schwinger Model. The bosonized form of the Gauss law can now be obtained from \eqref{eq:bos_action} as the equation of motion for $ A_0 $:
\begin{equation}
	\partial_{1}F_{01}=\frac{e^{2}}{\sqrt{\pi}}\partial_{1}\varphi
	\label{eq:BozGaussLaw}
\end{equation}
and the equation of motion for the boson becomes the equation of motion for a massive boson with mass $ m=\frac{e}{\sqrt{\pi}} $.

Before moving on to the solution of the Schwinger Model, we make two important comments. 
First, we find the expression for the electric charge in the bosonized form of the Schwinger Model. Once again using the duality \eqref{eq:bosonization_1}, we obtain
\begin{equation}
	Q=\int_{x_1}^{x_2}j_{0}dx=
	-\frac{1}{\sqrt{\pi}}\int_{x_1}^{x_2}\widetilde{k}_0 dx=
	\frac{1}{\sqrt{\pi}}\int_{x_1}^{x_2}\partial_{1}\varphi dx=
	\frac{1}{\sqrt{\pi}}\left( \varphi\left(x_2\right)-\varphi\left(x_1\right)\right)
	\label{eq:tot_charge}
\end{equation}

Second, we note that the boson is now identified up to $ R_\varphi=\sqrt{\pi} $, as opposed to $ R_{\varphi,free}=2\sqrt{\pi} $ in the free boson theory. This results from the fact that the translation $ \varphi\rightarrow\varphi+\sqrt{\pi} $ is now a gauge symmetry, and not a global symmetry. To show this, we start with the global symmetry group of the free fermion, which was shown to be $\left(U_A\left(1\right)\times U_V\left(1\right)\right)/\mathbb{Z}_2$ in section \ref{bosonization}. The Schwinger Model is obtained by gauging the vector symmetry $ U_V\left(1\right) $, and so the resulting global symmetry group of the Schwinger Model is $ U_A\left(1\right)/\mathbb{Z}_2 $. We conclude that the axial transformation by $ \pi $, acting as $ \psi\rightarrow e^{i\pi\gamma_5}\psi=-\psi $, is no longer a global symmetry, but a gauge symmetry\footnote{A more naive explanation would simply be that the axial transformation by $ \pi $ is identical to a vector transformation by $ \pi $, which is now a gauge transformation.}. In the bosonized theory, this corresponds to the global symmetry group being $ U_T\left(1\right)/\mathbb{Z}_2 $, with $ \varphi\rightarrow\varphi+\sqrt{\pi} $ being a gauge symmetry. This in turn leads to $ \varphi $ being identified up to $ R_\varphi=\sqrt{\pi} $, as required.



In order to solve the Schwinger Model, one must now eliminate the photon $ A_\mu $ from the action. This is always possible in $ 1+1 $ dimensions, since physical photons do not exist; photons are required to be transverse, while transverse directions do not exist in $ 1 $ spatial dimension. Since $ A_\mu $ contains two degrees of freedom (for $ \mu=0,1 $), eliminating it requires two constraints, which will be given by the gauge fixing condition and the Gauss law. Since this procedure depends on the topology of our space, we will discuss the solution in each topology separately in the following sections.

\subsection{Boundary Conditions}\label{boundary_conditions}

We are interested in the Schwinger Model on non-trivial topologies, and so we find the possible boundary conditions for the Schwinger Model. These are found by varying the action \eqref{eq:bos_action} and demanding that the boundary terms vanish. We find two possible boundary conditions for the gauge field:
\begin{subequations}\label{eq:gauge_bcs}
	\begin{gather}
	A_0=0
	\label{eq:gauge_bcs_1}\\
	F_{01}=\frac{e^2}{\sqrt{\pi}}\varphi
	\label{eq:gauge_bcs_2}
	\end{gather}
\end{subequations}
And two possible boundary conditions for the boson:
\begin{subequations}\label{eq:boson_bcs}
	\begin{gather}
	\varphi=0\mod\sqrt{\pi}
	\label{eq:boson_bcs_1}\\
	\partial_x\varphi=0
	\label{eq:boson_bcs_2}
	\end{gather}
\end{subequations}
Note that since the boson is compact with $ R_{\varphi}=\sqrt{\pi} $, one must make sure to sum over all possible choices $ \varphi=n\sqrt{\pi} $ for $ n\in\mathbb{Z} $ when choosing boundary condition \eqref{eq:boson_bcs_1}.

Our objective is to find charged states, and our choice of boundary condition should reflect this fact. A glance at the expression for the total charge \eqref{eq:tot_charge} convinces us to focus on Dirichlet boundary conditions \eqref{eq:boson_bcs_1} for the boson. On a compact spatial dimension $ x\in\left[0,L\right] $, one can then choose the boundary conditions $ \varphi\left(0\right)=k\sqrt{\pi} $ and $ \varphi\left(L\right)=l\sqrt{\pi} $ (called a twisted sector), which will result in an overall charge $ Q=l-k $ in the system. As for the gauge field, both choices in \eqref{eq:gauge_bcs} are relevant, and we shall consider both in our analysis.

\section{The Schwinger Model on $\mathbb{R} \times \mathbb{R} $}\label{RxR}

We proceed to present the solution of the Schwinger Model on $ \mathbb{R} \times \mathbb{R}$, following \cite{Coleman1976}. Specifically, we assume that all fields vanish at infinity.

Having found the bosonized form of the Schwinger Model \eqref{eq:bos_action}, we are left with eliminating the gauge field $ A_\mu $ from the action. This is achieved using two constraints: the gauge fixing condition and the Gauss law.
On $ \mathbb{R} \times \mathbb{R}$, axial gauge $ A_1=0 $ is a valid gauge choice, thus removing one of the degrees of freedom\footnote{note that in $ 1+1 $ dimensions, axial gauge also implies Coulomb gauge, $ \partial_1A_1=0 $}.
The Gauss law \eqref{eq:BozGaussLaw} in axial gauge becomes 
\begin{equation}
\partial_{1}^{2}A_{0}=-\frac{e^{2}}{\sqrt{\pi}}\partial_{1}\varphi
\label{eq:gauss_law_R}
\end{equation}
Integrating the Gauss law \eqref{eq:gauss_law_R}, and assuming that all fields vanish at infinity, we obtain 
\begin{equation}
\partial_{1}A_{0}=-\frac{e^{2}}{\sqrt{\pi}}\varphi
\end{equation}
In order to satisfy this constraint, we plug it into the bosonized action \eqref{eq:bos_action} to obtain:
\begin{equation}
S=\int d^{2}x\left[\frac{1}{2}\left(\partial_{\mu}\varphi\right)^{2}-\frac{1}{2}\frac{e^{2}}{\pi}\varphi^{2}\right]
\label{eq:mas_boson}
\end{equation}
This is the action of a free massive boson, with mass $m = \frac{e}{\sqrt\pi}$. 
We have thus shown that the Schwinger Model on $ \mathbb{R}\times\mathbb{R} $ is dual to a free massive boson. 
This boson is interpreted as an electron-positron pair connected by a flux tube, with the mass corresponding to the energy of the flux tube. Since the spectrum consists only of this neutral boson, we conclude that the theory does not permit charged states in free space, which proves that it is indeed confining.

We note that the action we have obtained in equation \eqref{eq:mas_boson} is inconsistent with the identification $ \varphi\sim\varphi+\sqrt{\pi} $ discussed in section \ref{bos_of_schwinger}, which hints that the naive treatment above isn't complete. Specifically, demanding that all fields vanish at infinity is too strong a contraint, and instead one must consider the more general boundary condition \eqref{eq:boson_bcs_1}, which will be discussed in the following sections.

\section{The Schwinger Model on $\mathbb{R} \times S^1 $}\label{S1}

We proceed to present the solution of the Schwinger Model on $ \mathbb{R} \times S^1$, where the circle has length $ L $. This analysis follows Manton \cite{Manton:1985}, save for the discussion regarding the zero modes. Our starting point is once again the bosonized form of the Schwinger Model \eqref{eq:bos_action}.

The analysis in section \ref{boundary_conditions} might lead us to expect twisted sectors in the theory, and hence the appearance of charged states. However, the Gauss law squanders our hopes. Integrating the Gauss law \eqref{eq:BozGaussLaw} gives
\begin{equation}
F_{01}\left(L\right) - F_{01}\left(0\right) = \frac{e^2}{\sqrt{\pi}}\left[\varphi\left(L\right) - \varphi\left(0\right)\right]
\end{equation}
We conclude that a twisted sector will impose $ F_{01}\left(0\right)\neq F_{01}\left(L\right) $. However, $ F_{01} $ is the electric field, and thus is a physical observable, which must be periodic on the circle $ S^1 $. Twisted sectors are thus forbidden, and so we may only consider boundary conditions for which $ \varphi\left(0\right)=\varphi\left(L\right) $. In particular, no total charge can appear in the system.

We must also choose a boundary condition for the gauge field. We shall consider only Dirichlet boundary conditions \eqref{eq:gauge_bcs_1}, since Neumann boundary conditions \eqref{eq:boson_bcs_2} result in a simpler theory. This results from the fact that the combination of the Gauss law \eqref{eq:BozGaussLaw} and the  boundary condition \eqref{eq:boson_bcs_2} impose the constraint $ F_{01}=\frac{e^2}{\sqrt{\pi}}\varphi $ on the theory, resulting in the action \eqref{eq:bos_action} reducing to that of a massive boson on a circle with periodic boundary conditions.

We can now proceed as before in order to eliminate the photon from the action.

\subsubsection*{Gauge Fixing Condition}

On $ S^{1} $, axial gauge is no longer a consistent gauge choice. This can be seen by considering the Wilson lines of the theory. Specifically, we consider the Wilson line $ P=\exp\left(i\int_{0}^{L} dxA_1\right)$. Due to the topology of $ S^1 $, this Wilson line becomes a Wilson loop, meaning that it is gauge invariant. However, choosing axial gauge sets $ P=1 $, whereas generically $ P\ne1 $\footnote{For example, taking $ A_1=Et $, representing a constant electric field in the x direction.}. Since a gauge invariant object cannot be modified by a choice of gauge fixing condition, we conclude that axial gauge is forbidden on $ \mathbb{R}\times S^1 $. 

Instead, we choose to work with Coulomb gauge, $ \partial_{1}A_{1}=0 $. One can show that this is a good gauge choice by directly defining the appropriate gauge transformation $ \Omega\left(x\right) $. However, this gauge choice leaves us with a residual gauge symmetry given by "large" gauge transformations, which act on $ A_1 $ by $ A_{1}\rightarrow A_{1}+\frac{2\pi}{L}n $ for $ n\in\mathbb{Z} $. Indeed, a general Wilson loop is invariant under these transformations. This results in $ A_1 $ being indentified up to $ \frac{2\pi}{L} $.

To summarize, our gauge fixing condition results in $ A_1 $ being spatially constant with $ A_1 \thicksim A_1 + \frac{2\pi}{L}$.

\subsubsection*{Gauss Law}

Having fixed the gauge, we are left with imposing the Gauss law on our theory. In Coulomb gauge, the Gauss law is once again
\begin{equation}
\partial_{1}^{2}A_{0}=-\frac{e^{2}}{\sqrt{\pi}}\partial_{1}\varphi
\label{eq:gauss_law_S1}
\end{equation}
In order to solve the Gauss law, we move to Fourier space. Since twisted boundary conditions are forbidden, $ \varphi $ is periodic and so the expansion of $ \varphi $ is given by
\begin{equation}
\varphi=
\varphi_0\left(t\right)
+\sqrt{\frac{2}{L}}\sum_{p\in\frac{n\pi}{L}}\varphi_{p}\left(t\right)\sin\left(px\right)
\end{equation}
and similarly for $ A_0,A_1 $ (note that $ A_1 $ is spatially constant, and so its expansion consists only of a zero mode $ A_{1,p=0} $).
The Gauss law imposes
\begin{equation}
A_{0,p}=-\frac{e^{2}}{i\sqrt{\pi}p}\varphi_{p},\;\;\;\;\;\;p\neq0
\label{eq:A0_from_gauss}
\end{equation}
which can be used to directly eliminate $ A_{0} $ from the action.

\subsection{Solution}

The gauge fixing condition and the Gauss law can now be used in order to remove $ A_\mu $ from the action. Plugging the Gauss law \eqref{eq:A0_from_gauss} and the Fourier expansions into the bosonized action \eqref{eq:bos_action}, we obtain:
\begin{equation}
S=\int dt\left[
\frac{1}{2}\left(\partial_{t}\varphi_0\right)^{2}
+\frac{1}{2e^{2}}\left(\partial_{t}A_{1,p=0}\right)^{2}
-\frac{1}{\sqrt{\pi}}\partial_{t}A_{1,p=0}\varphi_0+
\frac{1}{2}\sum_{p\neq0}\left(\left(\partial_{t}\varphi_p\right)^{2}
-\left(p^{2}+\frac{e^{2}}{\pi}\right)\varphi_p^{2}\right)
\right]
\end{equation}
The resulting Hamiltonian is:
\begin{equation}
\mathcal{H}=
\frac{1}{2}\left(e\pi_{A}+\frac{e}{\sqrt{\pi}}\varphi_0\right)^{2}+\frac{1}{2}\pi_{\varphi,0}^{2}+
\frac{1}{2}\sum_{p\neq0}\left(\pi^{2}_{\varphi,p}+\left(p^{2}+\frac{e^{2}}{\pi}\right)\varphi_p^{2}\right)
\label{eq:S1_hamiltonian}
\end{equation}
where $ \pi_{A} $ is the momentum conjugate to $ A_{1,p=0} $ and $ \pi_{\varphi,p} $ is the momentum conjugate to $ \varphi_p $. We find that the Hamiltonian for the non-zero modes $ p\neq0 $ is simply that of a free massive boson. The Hamiltonian for the zero mode $ p=0 $ is slightly more interesting, and we now proceed to analyze it further.

In order to put the Hamiltonian for the zero modes $ \mathcal{H}_0 $ into a more familiar form, we define new coordinates by
\begin{equation}
x=\sqrt{L}\varphi_0,\qquad y=\frac{\sqrt{L}}{e}A_{1,p=0} 
\end{equation}
Note that since both $ \varphi $ and $ A_1 $ are compact, our new coordinates are periodic, with $ L_x=\sqrt{\pi}L $ and $ L_y=\frac{2\pi}{L} $, so that they define a 2 dimensional torus. In terms of our new coordinates, the Hamiltonian becomes
\begin{equation}
\mathcal{H}_{0} = \frac{L}{2}p_x^{2} + \frac{L}{2}\left(p_y + \frac{e}{\sqrt{\pi}L}x\right)^{2}
\label{eq:zero_mode_hamiltonian}
\end{equation}
We recognize \eqref{eq:zero_mode_hamiltonian} as the Hamiltonian for a particle with mass $ M=\frac{1}{L} $ in the presence of a magnetic field $ \overline{B}=\frac{e}{\sqrt{\pi}L}\hat{z} $ on a 2d torus. 

The solution to this problem \cite{AlHashimi} is a slight variation of the well-known Landau Levels, which are the solutions of the same problem on the 2 dimensional plane. The most prominent effect of considering the problem on a torus instead of the plane is a reduction in the degeneracy of the energy levels; while the Landau Levels are infinitely degenerate, on the torus $ L_x\times L_y $ the degeneracy of each energy level is given by $ N=\frac{L_{x}L_{y}B}{2\pi}$. The energy levels themselves are unchanged, and are given by
$ E_n=\frac{B}{M}\left(n+\frac{1}{2}\right) $ for $ n\in\mathbb{N} $. Plugging in the relevant values for $ M $ and $ B $ we find that the zero modes have energy levels 
\begin{equation} 
E_n = \frac{e}{\sqrt{\pi}} \left(n+\frac{1}{2}\right), \qquad n\in\mathbb{N} 
\end{equation}
with $ N=1 $, meaning that they are not degenerate.

\subsection{Summary}

We have found that the Hamiltonian of the Schwinger model on $S^{1}$ with  Dirichlet boundary conditions for $ A_\mu $ and periodic boundary conditions for $ \varphi $ is given by equation \eqref{eq:S1_hamiltonian}. The spectrum of the model consists of two parts:
\begin{enumerate}
	\item The modes for $p\neq0$ are those of a massive boson with mass $m=\frac{e}{\sqrt{\pi}}$.
	\item The zero mode behaves like a QM particle in a constant magnetic field
	on a torus, with non-degenerate energy levels $\frac{e}{\sqrt{\pi}}\left(n+\frac{1}{2}\right)$ for $n\in\mathbb{N}$.
\end{enumerate}

\section{The Schwinger Model on $ \mathbb{R}\times \left[0,L\right] $}\label{RxI}

We now turn to the more difficult problem of the Schwinger Model on the interval $ \left[0,L\right] $. Since this problem is much more challenging, we will not attempt to solve it exactly, and instead we focus on finding charged boundary states and calculating their energy. 

\subsubsection*{Boundary Conditions}

The discussion in section \ref{boundary_conditions} convinces us to consider Dirichlet boundary conditions \eqref{eq:boson_bcs_1} for the boson $ \varphi $ at both ends of the interval, resulting in the total charge of the system being non-zero on a twisted sector.

As for the gauge boson, we shall consider both Dirichlet boundary conditions \eqref{eq:gauge_bcs_1} and Neumann boundary conditions \eqref{eq:gauge_bcs_2} in the following. For simplicity, we choose the same boundary condition at both ends of the interval.

\subsubsection*{Gauge Choice}
Once again, we choose Coulomb gauge, $ \partial_1A_1=0 $. We note that the discussion in section \ref{S1} regarding the Wilson Loop is irrelevant on the interval, since the Wilson Line $ P=e^{\int_0^L dxA_1} $ is no longer gauge invariant. Instead, it transforms under gauge transformations as $ P\rightarrow e^{iq}P $ for some constant $ q $. We conclude that Wilson Lines of the form $ P $ have charge $ q $ under the $ U\left(1\right) $ transformations. Specifically, we no longer have the residual gauge symmetry under which $ A_1\sim A_1+\frac{2\pi}{L} $.

\subsection{Non-relativistic limit}
As a warm-up, we consider the non-relativistic limit of this problem. Consider first the non-relativistic limit of the Schwinger model on $ \mathbb{R}\times\mathbb{R} $. We give the electrons a mass $ m $, and assume $ m\gg e $. In this limit, the Hamiltonian of an electron-positron pair connected by a flux tube is given by\footnote{This results from the fact that the energy of a flux tube with constant electric field $ E $ in the Schwinger Model is $ U=\frac{1}{2}\int E^2 dx = \frac{1}{2}E^2\ell $, where $ \ell $ is the length of the flux tube.}
\begin{equation*}
	H=\frac{1}{2m}p_1^2 + \frac{1}{2m}p_2^2 +  \frac{e^2}{2}\left|x_1-x_2\right|^2
\end{equation*}

Moving to center of mass coordinates (defined by $ X=\frac{x_1+x_2}{2} $, $ x=x_1-x_2 $) and ignoring the center of mass energy, we obtain the Hamiltonian

\begin{equation}
H=\frac{1}{2\mu}p^2 +\frac{e^2}{2}\left|x\right|^2\label{COMhamiltonian}
\end{equation}

where $ \mu=\frac{m}{2} $ is the reduced mass. Solving the Schrodinger equation, we find that the eigenfunctions of this Hamiltonian are the Airy functions, and the eigenenergies (assuming an antisymmetric wavefunction) are given by
\begin{equation*}
	E_i=\sqrt[3]{\frac{e^4}{4m}}\left(-a_i\right)
\end{equation*}
where $ a_i $ is the $ i $-th root of the Airy function.

We now solve the same problem on the half line $ [0,\infty) $. Our choice of boundary condition will be $ A_0(0)=0 $ (meaning that we place a conductor at the boundary), and we shall also demand that the wavefunction vanishes at the boundary. Placing an electron at a distance $ x $ from the boundary will create an image charge beyond the boundary, and the potential energy of such a configuration will be $ \frac{e^2}{2}x $. The Hamiltonian of a single electron at the boundary is thus given by

\begin{equation*}
	H=\frac{1}{2m}p^2 + \frac{e^2}{2}x^2
\end{equation*}

This Hamiltonian is identical to \eqref{COMhamiltonian}, apart from the mass of the particle. The eigenenergies of a single electron at the boundary are thus given by

\begin{equation*}
	E_i=\sqrt[3]{\frac{e^4}{8m}}\left(-a_i\right)
\end{equation*}

We have thus found charged particles near the boundary in the non-relativistic limit of the Schwinger Model.

\subsection{Solution for Dirichlet Boundary Conditions}\label{Dirichlet_bcs}

We now return to the (relativistic) Schwinger Model on an interval. We study the theory with Dirichlet boundary conditions for the gauge boson, given by \eqref{eq:gauge_bcs_1}. Explicitly, the boundary conditions for a general twisted sector are
\begin{subequations}\label{eq:twisted_bcs}
	\begin{gather}
	A_{0}\left(0\right)=A_{0}\left(L\right)=0
	\label{eq:twisted_bcs_1}\\
	\varphi\left(0\right)=k\sqrt{\pi},\quad
	\varphi\left(L\right)=l\sqrt{\pi}
	\label{eq:twisted_bcs_2}
	\end{gather}
\end{subequations}
for any $ k,l\in\mathbb{Z} $. Using the expression for the overall charge in the system \eqref{eq:tot_charge}, we find that the overall charge is $Q=l-k$. We  shall see that this can be interpreted as having $ k $ electrons at $x=0$, and $ l $ positrons at $x=L$ (where negative $ k $ means we have positrons instead of electrons, and similarly for negative $ l $).

Since this boundary condition forces the electric potential to vanish, we interpret it as placing conductors at both ends of the interval. We thus have a classical argument for the existence of charged boundary states near the boundary - an electron near the conductor would be part of a strongly bound electron-positron pair, with the positron being the image charge on the other side of the conductor.


We proceed to find the mode expansions of the fields. The equation of motion for $ \varphi $ is still that of a massive boson with mass $ m $, and so the mode expansion for $ \varphi $ with the boundary condition \eqref{eq:twisted_bcs_2} is:
\begin{equation}
\varphi=
\varphi_{0}(x)
+\sqrt{\frac{2}{L}}\sum_{p\in\frac{n\pi}{L}}\varphi_{p}\left(t\right)\sin\left(px\right)
\label{eq:BosModeExpansion}
\end{equation}
where $ \varphi_0 $ is given by
\begin{equation}
\varphi_{0}\left( x \right) =
\sqrt{\pi}\left[\frac{l-k\cosh\left(mL\right)}{\sinh\left(mL\right)}\sinh\left(mx\right)+k\cosh\left(mx\right)\right]
\label{eq:zero_mode}
\end{equation}

In order to understand the charged states, we must study $ \varphi_0 $. This is due to the fact that the charged states result from choosing twisted boundary conditions, and $ \varphi_0 $ is the only part of the mode expansion which contains any information about the boundary conditions. The shape of $ \varphi_0 $ strengthens our interpretation of the charged states being confined to the boundary; examining it for large $ L $, we see that it peaks at the boundaries, and decays exponentially away from each boundary, so that $ \varphi_{0}\approx0 $ in the bulk. Using equation \eqref{eq:tot_charge}, we thus conclude that the charge is confined to the boundaries, with charge $ -k $ at $ x=0 $ and charge $ l $ at $ x=L $. We thus interpret this result as having $ k $ electrons confined to the boundary at $ x=0 $, and $ l $ positrons confined to the boundary at $ x=L $.

Moving on, we once again use the Gauss law in order to obtain $A_{0}$. In Coulomb gauge, the Gauss law is once again given by
\begin{equation}
	\partial_{1}^{2}A_{0}=-\frac{e^{2}}{\sqrt{\pi}}\partial_{1}\varphi
\end{equation}
Solving for $ A_0 $, we obtain
\begin{equation}
A_{0}\left(x\right) = -\frac{e^{2}}{\sqrt{\pi}}\int_{0}^{x}\varphi_{0}\left(x^{\prime}\right)dx^{\prime} + \sqrt{\frac{2}{L\pi}}e^{2}\sum_{p\in\frac{n\pi}{L}}\frac{1}{p}\varphi_{p}\left(t\right)\cos\left(px\right) + C\left(t\right)x + D\left(t\right)
\end{equation}
where $ C,D $ are constants of integration. They are determined by demanding the boundary conditions, giving
\begin{equation}
A_{0}\left(x\right) = -\frac{e^{2}}{\sqrt{\pi}}\left(1+\frac{x}{L}\right)\int_{0}^{x}\varphi_{0}\left(x^{\prime}\right)dx^{\prime} + \sqrt{\frac{2}{L\pi}}e^2 \sum_{p\in\frac{n\pi}{L}}\frac{1}{p}\varphi_{p}\left(t\right) \left(\cos\left(px\right)-1\right) + \frac{2\sqrt{2}e^{2}}{\sqrt{\pi L}}\frac{x}{L}
\sum_{p\in\frac{\left(2n+1\right)\pi}{L}}\frac{1}{p}\varphi_{p}\left(t\right)
\end{equation}
Finally, our gauge choice once again fixes $ A_1=A_{1,p=0}\left(t\right) $. 

Plugging the expansions back into the action leads to the Hamiltonian splitting into even and odd modes, $\mathcal{H}=\mathcal{H}_{\mbox{even}}+\mathcal{H}_{\mbox{odd}}$,
with
\begin{equation}
\mathcal{H}_{\mbox{even}}=\frac{1}{2}\sum_{p\in\frac{2n\pi}{L}}\left(\pi_{\varphi,p}^{2}+\left(p^{2}+m^{2}\right)\varphi_{p}^{2}\right) + E_0
\end{equation}
and
\begin{equation}
\mathcal{H}_{\mbox{odd}}=\frac{e^{2}}{2}\pi_{A}^{2}+\frac{1}{2}\sum_{p\in\frac{\left(2n+1\right)\pi}{L}}\left(\pi_{\varphi,p}^{2}+\left(p^{2}+m^{2}\right)\varphi_{p}^{2}+\sqrt{\frac{2}{\pi}}\frac{2e^{2}}{pL}\pi_{A}\varphi_{p}\right)
\end{equation}

Where $ \pi_A $ is the momentum conjugate to $ A_{1,p=0} $, $ \pi_{\varphi,p} $ is the momentum conjugate to $ \varphi_p $, and $ E_0 $ is given by
\begin{equation}
	E_0=\frac{m\pi}{2}\left[\left(k^{2}+l^{2}\right)\coth\left(mL\right)
	-\frac{2kl}{\sinh\left(mL\right)}
	-\frac{\left(k+l\right)^{2}}{mL}\tanh^{2}\left(\frac{mL}{2}\right)\right]
	\label{eq:dirichlet_energy}
\end{equation}
We thus conclude that the even modes are simply massive boson modes, while the odd modes resemble a particle in a magnetic field.

We can now find the energy of the boundary states. This energy is given by the difference between the energy of a twisted sector with $ k,l\ne0 $ and that of the untwisted sector with $ k=l=0 $. Taking the difference of the corresponding Hamiltonians, we find that the energy of the boundary states is precisely $ E_0 $ defined in equation \eqref{eq:dirichlet_energy}, which we interpret as the energy associated with adding $ k $ electrons at $x=0$ and $ l $ positrons at $x=L$.

\subsection{Solution for Neumann Boundary Conditions}\label{Neumann_bcs}

We now study the theory with Neumann boundary conditions for the gauge boson \eqref{eq:gauge_bcs_2} on a general twisted sector. These are explicitly given by 
\begin{subequations}\label{eq:neumann_twisted_bcs}
	\begin{gather}
	\varphi\left(0\right)=k\sqrt{\pi},\qquad
	\varphi\left(L\right)=l\sqrt{\pi}
	\label{eq:neumann_twisted_bcs_2}\\
	F_{01}\left(0\right)=k e^2,\qquad F_{01}\left(L\right)=l e^2
	\label{eq:neumann_twisted_bcs_1}
	\end{gather}
\end{subequations}
for any $ k,l\in\mathbb{Z} $. Once again, we have a classical argument which leads us to expect charged states near the boundary. Since the electric field of a charged particle is constant, this boundary condition is equivalent to placing a charged particle just outside of the boundary. We then expect a particle with the opposite charge to be attracted to it from inside the interval, confining it to the boundary. 

The calculation for Neumann boundary conditions is simpler than the calculation for Dirichlet boundary conditions done in the previous section, and so we omit most of the details.

The mode expansion for $ \varphi $ has already been obtained, and is given by \eqref{eq:BosModeExpansion}. 
We use the Gauss law \eqref{eq:BozGaussLaw} in order to obtain $A_{0}$:
\[
A_{0}\left(x\right)=-\frac{e^{2}}{\sqrt{\pi}}\int_{0}^{x}\varphi_{0}\left(x^{\prime}\right)dx^{\prime}+
\sqrt{\frac{2}{L}}\frac{e^{2}}{\sqrt{\pi}}\sum_{p\in\frac{n\pi}{L}}\frac{1}{p}\varphi_{p}\left(t\right)\cos\left(px\right)+Cx+D
\]
where $ C,D $ are once again obtained by imposing the boundary conditions. Only $ C $ is relevant in the following (since $ D $ does not appear in the action), with the boundary conditions imposing $ C=\partial_0 A_1 $. 

As before, we are interested in the difference between the Hamiltonian of the twisted and untwisted sectors, which will give us the energy associated with adding the charges at the boundary. The result is
\begin{equation}
E_0=\frac{\pi  m}{2} \left[ \left(k^2+l^2\right) \coth (L m)-\frac{2 k l}{\sinh\left(mL\right)}\right]
\label{eq:neumann_energy}
\end{equation}

\subsection{Discussion}
We analyze equations \eqref{eq:dirichlet_energy},\eqref{eq:neumann_energy} for the energies with Dirichlet and Neumann boundary conditions respectively. First, as a consistency check, we place charged particles on only one of the boundaries (that is, we set $ l=0 $) and we take the limit $ eL\ll 1 $, which should correspond to the free theory result. The result in both \eqref{eq:dirichlet_energy},\eqref{eq:neumann_energy} is
\begin{equation}
E_0(mL\rightarrow0)=\frac{\pi  k^2}{2L}
\label{eq:energy_at_ml_gt_0}
\end{equation}
Which agrees with the free theory for the bosonic zero modes.

We can also consider the next order term in $ eL $. This term should correspond to the first order correction when introducing an electric field to the problem. Since introducing an electric field attaches an electric flux to each particle, we would expect the first order correction to result from the energy of this flux tube, which is proportional to $ e^2\ell $ where $ \ell $ is the length of the flux tube. Thus, we would expect the first order correction to be proportional to $ e^2 L $. We find that for small $ eL $, the first order correction to the energy for Dirichlet boundary conditions is $ \frac{k^2}{24}e^2L $, while for Neumann boundary conditions it is $ \frac{k^2}{6}e^2L $, and so both agree with our expectations.

Next, in order to better understand our result,  we take the limit of $ eL\gg1 $ (so that we are now considering the theory on the half-line $ \left[0,\infty\right) $). This will simplify the problem since it allows us to ignore effects resulting from interactions between the two boundaries. In this limit, both \eqref{eq:dirichlet_energy} and \eqref{eq:neumann_energy} reduce to
\begin{equation}
E_0\left(eL\rightarrow\infty \right) =
\frac{m\pi}{2}\left(k^{2}+l^{2} \right) \label{eq:energy_L_inf}
\end{equation}
In particular, we conclude that in this limit the contributions to the energy from the two boundaries are independent. It thus suffices to only consider boundary states at $ x=0 $, meaning we can set $ l=0 $. The energy of $ k $ electrons at a boundary for both choices of boundary conditions is thus given by
\begin{equation}
E_0=\frac{\pi m}{2}k
^{2}=\frac{e\sqrt{\pi}}{2}k^{2}
\label{eq:energy_of_k_electrons}
\end{equation}

We can call this the mass of the boundary states.

We comment on two interesting aspects of equation \eqref{eq:energy_of_k_electrons}. First, we explain why the result is identical for both Dirichlet and Neumann boundary conditions. As explained in sections \ref{Dirichlet_bcs} and \ref{Neumann_bcs}, both boundary conditions can effectively be understood as pairing each boundary particle with an image antiparticle beyond the boundary. In both cases we thus have half of a flux tube connecting an electron-positron pair, meaning that we expect the energy to be identical for both setups. The only reason why the full expressions for the energies \eqref{eq:dirichlet_energy}, \eqref{eq:neumann_energy} aren't identical is interactions between the boundaries.

Second, we explain why the energy \eqref{eq:energy_of_k_electrons} depends on the square of the number of particles $ k^2 $. This too can be understood in terms of flux tubes. Due to the Pauli exclusion principle, we cannot place all $ k $ particles very close to the boundary, and instead each additional particle will be placed farther and farther away from the boundary. Assuming a typical displacement $ \ell\sim\frac{1}{e} $ between adjacent particles, the $ j $-th particle will be at a typical distance of $ j\cdot \ell $ from the boundary. The energy of its flux tube is then given by  $$U_j=\frac{1}{2}\int E^2 \sim\frac{1}{2}e^2\cdot j\ell $$The total energy will then be
\begin{equation*}
	E=\sum_{j=1}^{k}U_j\sim\sum_{j=1}^{k}\frac{1}{2}ej\sim e k^2
\end{equation*}
which explains the origin of the $ k^2 $ term.

To summarize, we have found charged boundary states in the Schwinger Model on the interval for both Dirichlet and Neumann boundary conditions for the gauge field. These states are interpreted as single electron (positron) states, which are part of an electron-positron pair with an image positron (electron) on the other side of the boundary. In particular, in the limit $ eL\rightarrow\infty $ the energy of $ k $ electrons near the boundary is given by equation \eqref{eq:energy_of_k_electrons}.
The existence of these charged states in a confining theory is a very interesting result, and could lead one to expect a similar result in QCD, where single quarks could be isolated near boundaries.

\bibliographystyle{utphys}
\providecommand{\href}[2]{#2}\begingroup\raggedright\endgroup

\end{document}